\documentclass[a4paper,11pt]{article}
\usepackage[sort&compress,numbers]{natbib}
\usepackage{pos}

\def\beqn{\begin{eqnarray}} \def\eeqn{\end{eqnarray}}
\def\beq{\begin{equation}} \def\eeq{\end{equation}}

\title{Reconstructing parton collisions with machine learning techniques}
\ShortTitle{Reconstructing parton collisions with machine learning techniques}
\author*[a,b]{G. F. R. Sborlini}
\author[c]{D. F. Renter\'ia-Estrada}
\author[c]{R. J. Hern\'andez-Pinto}
\author[d]{P. Zurita}

\affiliation[a]{Departamento de F\'isica Fundamental e IUFFyM, Universidad de Salamanca, E-37008 Salamanca, Spain}
\affiliation[b]{Escuela de Ciencias, Ingenier\'ia y Diseño, Universidad Europea de Valencia, Paseo de la Alameda 7, 46010 Valencia, Spain}
\affiliation[c]{Facultad de Ciencias F\'isico-Matem\'aticas, Universidad Aut\'onoma de Sinaloa, Ciudad Universitaria, CP 80000, Culiac\'an, Sinaloa, M\'exico}
\affiliation[d]{Institut f\"ur Theoretische Physik, Universit\"at Regensburg, 93040 Regensburg, Germany}

\emailAdd{german.sborlini@desy.de}

\abstract{Having access to the parton-level kinematics is important for understanding the internal dynamics of particle collisions. Here, we present new results aiming to an efficient reconstruction of parton collisions using machine-learning techniques. By simulating the collider events, we related experimentally-accessible quantities with the momentum fractions of the involved partons. We used photon-hadron production to exploit the cleanliness of the photon signal, including up to NLO QCD-QED corrections. Neural networks led to an outstanding reconstruction efficiency, suggesting a powerful strategy for unveiling the behaviour of the fundamental bricks of matter in high-energy collisions.}

\FullConference{
41st International Conference on High Energy physics - ICHEP2022\\
  6-13 July, 2022\\
  Bologna, Italy}

\begin{document}
\maketitle

%%%%%%%%%%%%%%%%%%%%%%%%%%%%%%%%%%%%%%%%%%%%%%%%%%%%
\section{Introduction and motivation}
\label{sec:introduction}
%%%%%%%%%%%%%%%%%%%%%%%%%%%%%%%%%%%%%%%%%%%%%%%%%%%%
In current high-energy experiments, bunches of particles collide, allowing their fundamental constituents to interact and produce new particles. The detectors identify the product of the collisions, but it is not straightforward to understand what is really going on during the process. In our recent work \cite{Renteria-Estrada:2021zrd}, we explore the application of Machine-Learning (ML) techniques to model the partonic momentum fractions in terms of experimentally-accessible variables. In first place, we calculated the differential hadronic cross-section for photon-hadron production including up to Next-to-Leading Order (NLO) QCD and Leading-Order (LO) QED corrections. Using a code based on Monte-Carlo (MC) integration \cite{deFlorian:2010vy,Renteria-Estrada:2021rqp}, we simulated the collisions and analyzed the events to determine the correlations among measurable and partonic quantities. Then, we applied ML algorithms that allow us to find the momentum fractions of the partons involved in the process in terms of suitable combinations of the final state momenta.

We tested our proof-of-concept (PoC) with photon-hadron production at colliders because the photon provides a clean probe to access the parton kinematics. The aim was to reconstruct the momentum fraction $x$ and $z$ of the partons coming from the protons and undergoing the fragmentation into a hadron, respectively. Due to the fact that parton kinematics are not physically-defined (i.e. it is a model), we provided a quantitative estimation of their most probable values. For this purpose, we trained Neural Networks (NN) to predict the MC partonic momentum fractions in terms of external momenta.

%%%%%%%%%%%%%%%%%%%%%%%%%%%%%%%%%%%%%%%%%%%%%%%%%%%%
\section{Computational setup}
\label{sec:Setup}
%%%%%%%%%%%%%%%%%%%%%%%%%%%%%%%%%%%%%%%%%%%%%%%%%%%%
In order to carry out the calculations, we relied on the well-known factorization theorem, which implies that the cross-section is described by the convolution between PDFs, FFs and the partonic cross-section. On top of that, we took advantage of the smooth-cone isolation \cite{Frixione:1998jh} with the purpose of avoiding the introduction of the photon fragmentation (whose accuracy is considerably lower than the other FFs). So, our starting point for the cross-section calculation was
\beq
d\sigma_{H_1 H_2  \to h \gamma} = \sum_{a_1 a_2 a_3} \int\, dx_1 \, dx_2 \, dz \, f_{H_1/a_1}(x_1,\mu_I) \, f_{H_2/a_2}(x_2,\mu_I) \, D_{a_3/h}(z,\mu_F) \, d\hat\sigma_{a_1\, a_2 \to a_3 \gamma}  \, ,
\label{eq:CrossSection}
\eeq
where we included up to NLO QCD and LO QED corrections. For the NLO part, we applied the FKS method \cite{Frixione:1995ms} to cancel the infrared singularities. %UV singularities were directly subtracted from the virtual corrections by using $\overline{{\rm MS}}$ renormalization. 
Once the IR-finite NLO differential cross-section was defined, we established a \emph{binning strategy} to discretize the phase-space. In this way, we managed to combine the real (2-to-3), virtual (2-to-2) and counter-terms (2-to-2) contributions by defining bins in the measurable variables, i.e. $\bar{{\cal V}}_{\rm Exp.} = \{{\bar p}_T^\gamma,{\bar p}_T^\pi,{\bar\eta}^\gamma,{\bar\eta}^\pi,{\overline{\cos}}(\phi^\pi-\phi^\gamma)\}$. Then, we integrated the fully differential cross-section from Eq. (\ref{eq:CrossSection}) in each bin, obtaining $\sigma_j$ for $p_j \in \bar{{\cal V}}_{\rm Exp.}$. This step was crucial, since the higher-order corrections to the cross-sections involve contributions living in different phase-spaces.

%%%%%%%%%%%%%%%%%%%%%%%%%%%%%%%%%%%%%%%%%%%%%%%%%%%%
\section{Parton kinematics reconstruction}
\label{sec:Reconstruction}
%%%%%%%%%%%%%%%%%%%%%%%%%%%%%%%%%%%%%%%%%%%%%%%%%%%%
For the process that we considered, i.e. $p+p \to \pi+\gamma$, the partonic momentum fractions $x_1$, $x_2$ and $z$ are unambiguously fixed by external particle’s kinematics only at LO. When computing higher-order corrections, real radiation including extra particles must be included and momentum conservation equations differs from LO kinematics. In order to obtain an approximation, we relied on the LO expressions and we found a decent agreement with the MC momentum fractions, as reported in Ref. \cite{deFlorian:2010vy}. However, in our recent work \cite{Renteria-Estrada:2021zrd}, we decided to tackle the problem by using more flexible ML techniques. 

\begin{figure}[h!]
    \centering
    \includegraphics[width=0.75\textwidth]{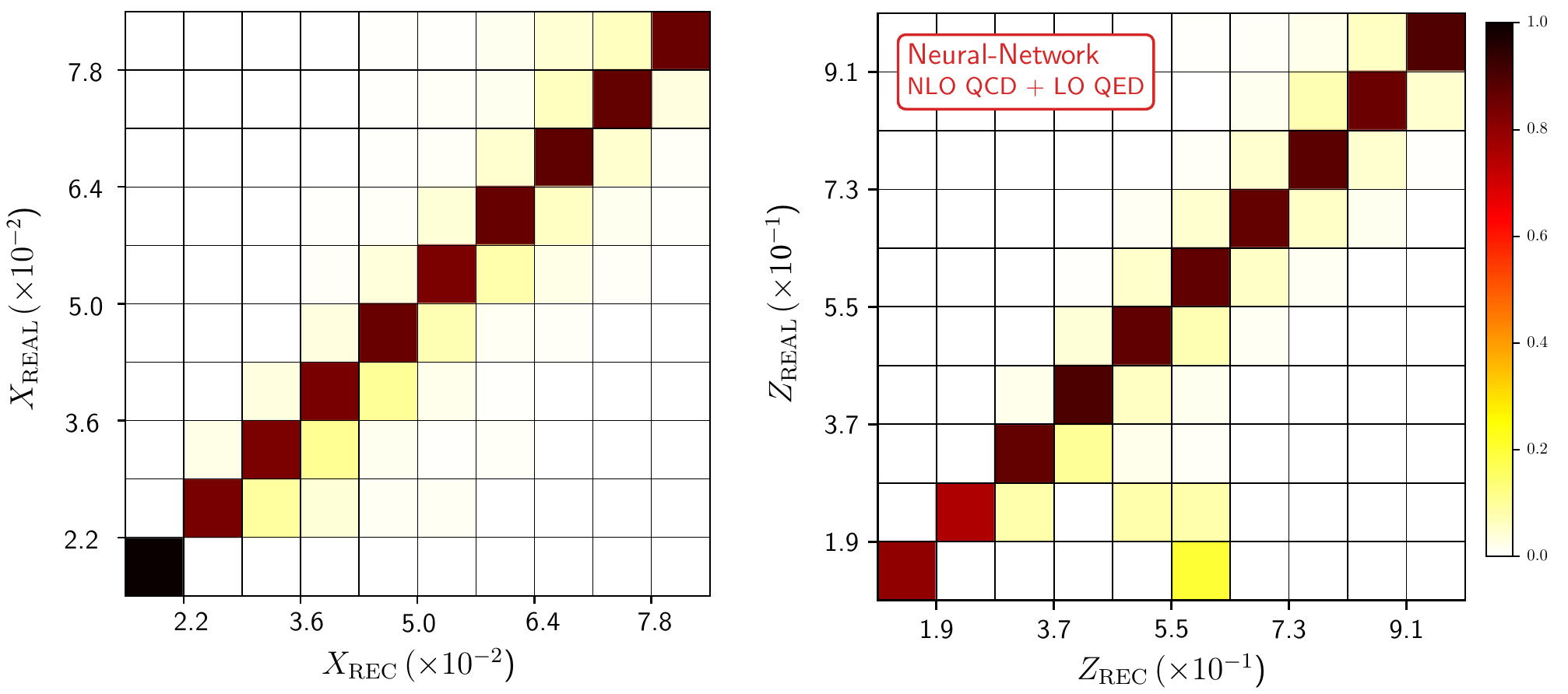}
    \caption{Comparison of the real momentum fractions ($y$-axis) and the reconstructed ones using Neural Networks ($x$-axis), for $x$ (left) and $z$ (right), respectively.}
    \label{fig:Figura1}
\end{figure}

The first step was to introduce the concept of \emph{effective momentum fractions} at higher-orders. Given a point in the grid, $p_j = \{ {\bar p}_T^\gamma,{\bar p}_T^\pi,{\bar\eta}^\gamma,{\bar\eta}^\pi,{\overline{\cos}}(\phi^\pi-\phi^\gamma) \} \in {\bar{\cal V}}_{\rm Exp}$, we defined 
\beq
(x_{1,2})_j = \sum_i \, (x_{1,2})_i \frac{d \sigma_j}{d x_{1,2}} (p_j;(x_{1,2})_i) \, , \quad  (z)_j = \sum_i \, z_i \frac{d \sigma_j}{d z} (p_j;z_i) \, ,
\label{eq:Averages}
\eeq
with the purpose of identifying a mean value for the MC momentum fractions associated to the sum of all the real-emission topologies contributing to the same measurable final-state configuration (i.e. the same point $p_j$). Then, we used ML to find mappings connecting the points in the grid with the average value of the momentum fraction per bin, i.e.   
\beq
X_{\rm REC} := {\bar{\cal V}}_{\rm Exp} \longrightarrow \bar{X}_{REAL} = \{(x)_j\} \, , \quad  Z_{\rm REC} := {\bar{\cal V}}_{\rm Exp} \longrightarrow \bar{Z}_{REAL} = \{(z)_j\} \, .
\label{eq:Mappings}
\eeq
We explored different approaches, being NN among them. The flexibility of the NN provided the best reconstruction with less constraints, as shown in the correlation plot in Fig. \ref{fig:Figura1}.   

\begin{figure}[h!]
    \centering
    \includegraphics[width=0.8\textwidth]{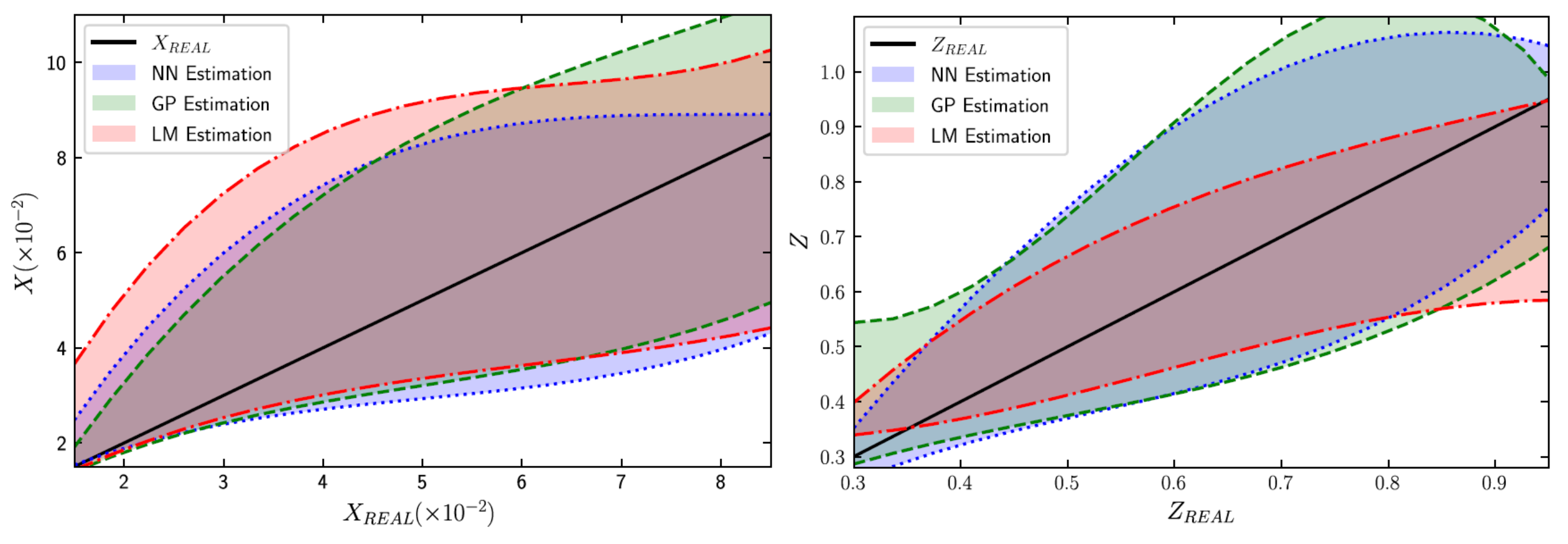}
    \caption{Correlation plots for $x_1$ (left) and $z$ (right), with the associated error bands due to the propagation of the scale uncertainties in the reconstruction. The Linear Method (LM), Gaussian Regression (GP) and Neural Networks (NN) were considered.}
    \label{fig:Figura2}
\end{figure}

Finally, we studied the propagation of the scale uncertainties by defining different datasets, varying the renormalization and factorization scales by a factor 2 up and down. With these datasets, we trained estimators (i.e. $X_{\rm REC}^{(\xi)}$, with $\xi$ the energy scale) and evaluated them in each point of the grid. As reported in Ref.\cite{Renteria-Estrada:2021zrd}, the average reconstruction error turned out to be $7 \%$ for $x$ and $5 \%$ for $z$. In Fig. \ref{fig:Figura2}, we show the error bands in the correlation plots, for different ML approaches. The bands are narrower for smaller values of $x$ and $z$, since the cross-section is higher and the training is more accurate.

%%%%%%%%%%%%%%%%%%%%%%%%%%%%%%%%%%%%%%%%%%%%%%%%%%%%
\section{Conclusions}
\label{sec:Conclusions}
%%%%%%%%%%%%%%%%%%%%%%%%%%%%%%%%%%%%%%%%%%%%%%%%%%%%
In this work, we presented a proof-of-concept (PoC) to reconstruct parton-level kinematics by using ML techniques. The results are in agreement with previous findings \cite{deFlorian:2010vy}, but they required much less human intervention. In particular, NN did not even need to select a function basis (which is an important choice in other approaches, such as Gaussian Regression or the Linear Method). Our results indicate that this PoC could be applied to other processes, and could be used to impose stricter constraints for PDF/FF determination.

%%%%%%%%%%%%%%%%%%%%%%%%%%%%%%%%%%%%%%%%%%%%%%%%%%%%
%%%%%%%%%%%%%%%%%%%%%%%%%%%%%%%%%%%%%%%%%%%%%%%%%%%%
\section*{Acknowledgments}
GS is supported by Programas Propios II (Universidad de Salamanca), EU Horizon 2020 research and innovation program STRONG-2020 project under grant agreement No. 824093 and H2020-MSCA-COFUND-2020 USAL4EXCELLENCE-PROOPI-391 project under grant agreement No 101034371. D. F. R.-E. and R. J. H.-P. are supported by CONACyT (Mexico) through the Project No. 320856 (Paradigmas y Controversias de la Ciencia 2022), Ciencia de Frontera 2021-2024 and PROFAPI 2022 Grant No. PRO-A1-024 (Universidad Autónoma de Sinaloa). P.Z. acknowledges support from the Deutsche Forschungsgemeinschaft (DFG, German Research Foundation)- Research Unit FOR 2926, grant number 430915485.
%%%%%%%%%%%%%%%%%%%%%%%%%%%%%%%%%%%%%%%%%%%%%%%%%%%%
%%%%%%%%%%%%%%%%%%%%%%%%%%%%%%%%%%%%%%%%%%%%%%%%%%%%

%\begin{normalsize}
%\bibliography{ref}
%\end{normalsize}

\providecommand{\href}[2]{#2}\begingroup\raggedright\endgroup

\end{document}